\documentclass[11pt]{article}
\usepackage[margin=2cm,left=3cm,right=3cm]{geometry}

\usepackage{amsmath,amsthm,amsfonts}
\usepackage{xcolor}
\usepackage{graphicx}
\usepackage{nicefrac}
\usepackage{xspace}
\usepackage{booktabs}
\usepackage{caption}
\usepackage{subcaption}
\usepackage{aligned-overset}
\usepackage{enumitem}
\usepackage{cite}
\usepackage{tikz}
\usetikzlibrary{automata, positioning}

\DeclareMathOperator{\BF}{BF}
\DeclareMathOperator{\OPT}{OPT}
\DeclareMathOperator{\E}{E}

\newcommand{\A}{\ensuremath{\mathcal{A}}}
\newcommand{\ALG}{\mathit{ALG}}
\newcommand{\eps}{\varepsilon}

\newcommand{\bestfit}{Best Fit\xspace}

\newtheorem{theorem}{Theorem}
\numberwithin{theorem}{section}
\newtheorem{lemma}[theorem]{Lemma}
\newtheorem{definition}[theorem]{Definition}

\newtheorem{proposition}[theorem]{Proposition}
\newtheorem{example}[theorem]{Example}
\newtheorem{claim}[theorem]{Claim}

\title{Best Fit Bin Packing with Random Order Revisited\thanks{Work supported by the European Research Council, Grant Agreement No.\ 691672. \\
		A preliminary version of this paper received the best paper award at \textit{45th International Symposium on
			Mathematical Foundations of Computer Science (MFCS 2020) and was published in the conference proceedings \cite{DBLP:conf/mfcs/Albers0L20}.}}
}

\author{Susanne Albers \\
		Department of Computer Science, \\Technial University of Munich,\\
		albers@in.tum.de \and 
		Arindam Khan\\
		Department of Computer Science and Automation, \\Indian Institute of Science,\\
		arindamkhan@iisc.ac.in \and 
		Leon Ladewig\\
		Department of Computer Science, \\Technial University of Munich,\\
		ladewig@in.tum.de
}

\date{}

\begin{document}

\maketitle

\begin{abstract}
	\bestfit is a well known online algorithm for the bin packing problem, where a collection of one-dimensional items has to be packed into a minimum number of unit-sized bins. In a seminal work, Kenyon [SODA 1996] introduced the (asymptotic) \textit{random order ratio} as an alternative performance measure for online algorithms. Here, an adversary specifies the items, but the order of arrival is drawn uniformly at random. Kenyon's result establishes lower and upper bounds of $1.08$ and $1.5$, respectively, for the random order ratio of \bestfit. Although this type of analysis model became increasingly popular in the field of online algorithms, no progress has been made for the \bestfit algorithm after the result of Kenyon.
	
	We study the random order ratio of \bestfit and tighten the long-standing gap by establishing an improved lower bound of $1.10$. For the case where all items are larger than $1/3$, we show that the random order ratio converges quickly to $1.25$.
	It is the existence of such large items that crucially determines the performance of \bestfit in the general case.
	Moreover, this case is closely related to the classical maximum-cardinality matching problem in the fully online model.
	As a side product, we show that \bestfit satisfies a monotonicity property on such instances, unlike in the general case.
	
	In addition, we initiate the study of the \textit{absolute} random order ratio for this problem. In contrast to asymptotic ratios, absolute ratios must hold even for instances that can be packed into a small number of bins.
	We show that the absolute random order ratio of \bestfit is at least $1.3$.
	For the case where all items are larger than $1/3$, we derive upper and lower bounds of $21/16$ and $1.2$, respectively.
\end{abstract}

\section{Introduction}
\label{sec:intro}

One of the fundamental problems in combinatorial optimization is \textit{bin packing}. 
Given a list $I=(x_1,\ldots,x_n)$ of $n$ items with sizes from $(0,1]$ and an infinite number of unit-sized bins, the goal is to pack all items into the minimum number of bins. Formally, a \textit{packing} is an assignment of items to bins such that for any bin, the sum of assigned items is at most 1.
While an offline algorithm has complete information about the items in advance, in the online variant, items are revealed one by one. Therefore, an online algorithm must pack $x_i$ without knowing future items $x_{i+1},\ldots,x_n$ and without modifying the packing of previous items.

As the problem is strongly $\mathsf{NP}$-complete \cite{DBLP:journals/jacm/GareyJ78}, research mainly focuses on efficient approximation algorithms. The offline problem is well understood and admits even approximation schemes \cite{DBLP:journals/combinatorica/VegaL81,DBLP:conf/focs/KarmarkarK82,DBLP:conf/soda/HobergR17}. 
The online variant is still a very active field in the community \cite{DBLP:journals/csr/ChristensenKPT17}, as the asymptotic approximation ratio of the best online algorithm is still unknown \cite{DBLP:conf/esa/BaloghBDEL18,DBLP:conf/waoa/BaloghBDEL19}.
%
As one of the first algorithms for the problem, Garey et al.\ proposed the algorithms \bestfit and First Fit \cite{DBLP:conf/stoc/GareyGU72}. Johnson published the Next Fit algorithm briefly afterwards
\cite{DBLP:journals/jcss/Johnson74}. All of these algorithms work in the online setting and attract by their simplicity: Suppose that $x_i$ is the current item to pack. The algorithms work as follows:
\begin{description}
	\item[\bestfit (BF)] Pack $x_i$ into the fullest bin possible, open a new bin if necessary.
	\item[First Fit (FF)] Maintain a list of bins ordered by the time at which they were opened. Pack $x_i$ into the first possible bin in this list, open a new bin if necessary.
	\item[Next Fit (NF)] Pack $x_i$ into the bin opened most recently if possible; open a new bin if necessary.
\end{description}
Another important branch of online algorithms is based on the \textsc{harmonic} algorithm \cite{DBLP:journals/jacm/LeeL85}. This approach has been massively tuned and generalized in a sequence of papers \cite{DBLP:journals/jal/RamananBLL89,DBLP:journals/jacm/Seiden02,DBLP:conf/esa/BaloghBDEL18}.

To measure the performance of an algorithm, different metrics exist. For an algorithm $\A$, 
let $\A(I)$ and $\OPT(I)$ denote the number of bins used by $\A$ and an optimal offline algorithm, respectively, to pack the items in $I$.  Let $\mathcal{I}$ denote the set of all item lists.
The most common metric for bin packing algorithms is the \textit{asymptotic (approximation) ratio} defined as
\begin{equation*}
R_{\mathcal{A}}^\infty = \limsup_{k \to \infty} ~ \sup_{I \in \mathcal{I}} \{ \mathcal{A}(I) / \OPT(I) \mid \OPT(I)=k \} \,.
\end{equation*}
Note that $R_{\mathcal{A}}^\infty$ focuses on instances where $\OPT(I)$ is large. 
This avoids anomalies typically occurring on lists that can be packed optimally into few bins.
However, many bin packing algorithms are also studied in terms of the stronger \textit{absolute (approximation) ratio}
\begin{equation*}
R_{\mathcal{A}} = \sup_{I \in \mathcal{I}} \{ \mathcal{A}(I) / \OPT(I) \} \,.
\end{equation*}
Here, the approximation ratio $R_{\mathcal{A}}$ must hold for each possible input. 
An online algorithm with (absolute or asymptotic) ratio $\alpha$ is also called \textit{$\alpha$-competitive}.

Table~{tab:relatedWork} shows the asymptotic and absolute approximation ratios of the three heuristics \bestfit, First Fit, and Next Fit. Interestingly, for these algorithms both metrics coincide.
While the asymptotic ratios of \bestfit and Next Fit were established already in early work \cite{DBLP:journals/siamcomp/JohnsonDUGG74}, the absolute ratios have been settled rather recently \cite{DBLP:conf/icalp/DosaS14,DBLP:conf/stacs/DosaS13}.

Note that the above performance measures are clearly worst-case orientated. An adversary can choose items and present them in an order that forces the algorithm into its worst possible behavior. In the case of \bestfit, hardness examples are typically based on lists where small items occur before large items \cite{DBLP:conf/stoc/GareyGU72}. In contrast, it is known that \bestfit performs significantly better if items appear in non-increasing order \cite{DBLP:journals/siamcomp/JohnsonDUGG74}.
For real-world instances, it seems overly pessimistic to assume adversarial order of input. Moreover, sometimes worst-case ratios hide interesting properties of algorithms that occur in average cases. This led to the development of alternative measures.

A natural approach that goes beyond worst-case was introduced by Kenyon \cite{DBLP:conf/soda/Kenyon96} in 1996. In the model of random order arrivals, the adversary can still specify the items, but the arrival order is permuted randomly. 
The performance measure described in \cite{DBLP:conf/soda/Kenyon96} is based on the asymptotic ratio, but can be applied to absolute ratios likewise. 
In the resulting performance metrics, an algorithm must satisfy its performance guarantee in expectation over all permutations. We define
\begin{align*}
RR_{\mathcal{A}}^\infty &= \limsup_{k \to \infty} ~ \sup_{I \in \mathcal{I}} \left\{ \E[\mathcal{A}(I^\sigma)] / \OPT(I) \mid \OPT(I)=k \right\} ~~~ \text{and} \nonumber \\
RR_{\mathcal{A}} &= \sup_{I \in \mathcal{I}} \left\{ \E[\mathcal{A}(I^\sigma)] / \OPT(I) \right\} 
\end{align*}
as the \textit{asymptotic random order ratio} and the \textit{absolute random order ratio} of algorithm $\A$, respectively.
Here, $\sigma$ is drawn uniformly at random from $\mathcal{S}_n$, the set of permutations of $n$ elements, and $I^\sigma = (x_{\sigma(1)},\ldots,x_{\sigma(n)})$ is the permuted list.

\begin{table}
	\caption{Approximation ratios in different metrics of common bin packing heuristics. In $R_\mathrm{NF}$, the symbol $\gamma$ refers to the total size of items in the list. }
	\label{tab:relatedWork}
	\centering
	\begin{tabular}{llll} \toprule
		Algorithm \A & Abs.\ ratio $R_\A$ & Asym.\ ratio $R_\A^\infty$ & Asym.\ random order ratio $RR_\A^\infty$ \\ \midrule
		\bestfit  & 1.7 \cite{DBLP:conf/icalp/DosaS14} & 1.7 \cite{DBLP:journals/siamcomp/JohnsonDUGG74} & $1.08 \leq RR_{BF}^\infty \leq 1.5$ \cite{DBLP:conf/soda/Kenyon96} \\
		First Fit & 1.7 \cite{DBLP:conf/stacs/DosaS13} & 1.7 \cite{DBLP:journals/siamcomp/JohnsonDUGG74} & ---  \\
		Next Fit  & $2- 1 / \lceil \gamma \rceil$ \cite{DBLP:journals/dam/BoyarDE12} & 2 \cite{DBLP:journals/jcss/Johnson74} & 2 \cite{DBLP:journals/dam/CoffmanCRZ08}  \\ \bottomrule
	\end{tabular}
\end{table}

\subsection{Related work}

The following literature review only covers results that are most relevant to our work. 
We refer the reader to the article \cite{coffman2013bin} by Coffman et al.\  for an extensive survey on (online) bin packing. For further problems studied in the random order model, see \cite{DBLP:journals/corr/abs-2002-12159}.

\paragraph{Bin packing.}

Kenyon introduced the notion of asymptotic random order ratio $RR_\A^\infty$ for online bin packing algorithms in \cite{DBLP:conf/soda/Kenyon96}. For the \bestfit algorithm, Kenyon proves an upper bound of 1.5 on $RR_\mathcal{\BF}^\infty$, demonstrating that random order significantly improves upon $R_{\BF}^\infty = 1.7$. However, it is conjectured in \cite{DBLP:conf/soda/Kenyon96,coffman2013bin} that the actual random order ratio is close to $1.15$.
The proof of the upper bound crucially relies on the following scaling property: With high probability, the first $t$ items of a random permutation can be packed optimally into $\frac{t}{n} \OPT(I) + o(n)$ bins.
On the other side, Kenyon proves that $RR_\mathcal{\BF}^\infty \geq 1.08$. This lower bound is obtained from the weaker i.i.d.-model, where item sizes are drawn independently and identically distributed according to a fixed probability distribution.

Coffman et al.\ \cite{DBLP:journals/dam/CoffmanCRZ08} analyzed next-fit in the random order model and showed that $RR_\mathcal{\mathrm{NF}}^\infty = 2$, matching the asymptotic approximation ratio $RR_{\mathrm{NF}}^\infty = 2$
(see Table~\ref{tab:relatedWork}). 
Fischer and R\"oglin \cite{DBLP:conf/latin/FischerR18} obtained analogous results for Worst Fit \cite{DBLP:journals/jcss/Johnson74} and Smart Next Fit \cite{DBLP:journals/ipl/Ramanan89}. Therefore, all three algorithms fail to perform better in the random order model than in the adversarial model.

A natural property of bin packing algorithms is monotonicity, which holds if an algorithm never uses fewer bins to pack $I'$ than for $I$, where $I'$ is obtained from $I$ by increasing item sizes.
Murgolo \cite{DBLP:journals/dam/Murgolo88} showed that next-fit is monotone, while \bestfit and First Fit are not monotone in general.
The concept of monotonicity also arises in related optimization problems, such as scheduling \cite{DBLP:journals/siamam/Graham69} and bin covering \cite{DBLP:conf/latin/FischerR18}.

\paragraph{Bin covering.}
The dual problem of bin packing is bin covering, where the goal is to cover as many bins as possible. A bin is covered if it receives items of total size at least 1.
Here, a well-studied and natural algorithm is Dual Next Fit (DNF).
In the adversarial setting, DNF has asymptotic ratio $R_\mathcal{\mathrm{DNF}}^\infty = 1/2$ which is best possible for any online algorithm \cite{DBLP:journals/tcs/ChristFL14}. Under random arrival order, Christ et al.\ \cite{DBLP:journals/tcs/ChristFL14} showed that $RR_{\mathrm{DNF}}^\infty \leq 4/5$. This upper bound was improved later by Fischer and R\"oglin \cite{DBLP:conf/latin/FischerR16} to $RR_{\mathrm{DNF}}^\infty \leq 2/3$. The same group of authors further showed that $RR_{\mathrm{DNF}}^\infty \geq 0.501$, i.e., DNF performs strictly better under random order than in the adversarial setting \cite{DBLP:conf/latin/FischerR18}.

\paragraph{Matching.} Online matching can be seen as the key problem in the field of online algorithms \cite{DBLP:journals/fttcs/Mehta13}. 
Inspired by the seminal work of Karp, Vazirani, and Vazirani \cite{DBLP:conf/stoc/KarpVV90}, who introduced the online bipartite matching problem with one-sided arrivals, the problem has been studied in many generalizations.
Extensions include 
fully online models \cite{DBLP:conf/focs/GamlathKMSW19,DBLP:conf/soda/0002PTTWZ19,DBLP:conf/stoc/0002KTWZZ18},
vertex-weighted versions \cite{DBLP:journals/talg/HuangTWZ19,ALHERZ2019}
and, most relevant to our work, random arrival order \cite{DBLP:journals/talg/HuangTWZ19,DBLP:conf/stoc/MahdianY11}.


\subsection{Our results}

While several natural algorithms fail to perform better in the random order model, \bestfit emerges as a strong candidate in this model. The existing gap between 1.08 and 1.5 clearly leaves room for improvement;
closing (or even narrowing) this gap has been reported as challenging and interesting open problem in several papers \cite{DBLP:journals/tcs/ChristFL14,DBLP:journals/dam/CoffmanCRZ08,DBLP:journals/corr/abs-2002-12159}.
To the best of our knowledge, our work provides the first new results on the problem since the seminal work by Kenyon.
Below we describe our results in detail. In the following theorems, the expectation is over the permutation $\sigma$ drawn uniformly at random.

If all items are strictly larger than $1/3$, the objective is to maximize the number of bins containing two items.
This problem is closely related to finding a maximum-cardinality matching in a vertex-weighted graph;
our setting corresponds with the fully online model studied in \cite{ALHERZ2019} under random order arrival.
Also in the analysis from \cite{DBLP:conf/soda/Kenyon96}, this special case arises. There, it is sufficient to argue that $\BF(I) \leq \frac{3}{2} \OPT(I) + 1$ under adversarial order. We show that \bestfit performs significantly better under random arrival order:

\begin{theorem}
	\label{theo:largeItemsUB}
	For any list $I$ of items larger than $1/3$, we have $\E[\BF(I^\sigma)] \leq \frac{5}{4} \OPT(I) + \frac{1}{4}$.
\end{theorem}

The proof of Theorem~\ref{theo:largeItemsUB} is developed in Section~\ref{sec:13largeItems} and based on several pillars. 
First, we show that \bestfit is monotone in this case (Proposition~\ref{prop:monotonicity}), unlike in the general case \cite{DBLP:journals/dam/Murgolo88}.
This property can be used to restrict the analysis to instances with well-structured optimal packing.
The main technical ingredient is introduced in Section~\ref{sec:13largeGoodOrder} with Lemma~\ref{lemma:goodOrderPairsLMbins} as the key lemma.
Here, we show that \bestfit maintains some parts of the optimal packing, depending on certain structures of the input sequence. We identify these structures and show that they occur with constant probability for a random permutation.
It seems likely that this property can be used in a similar form to improve the bound $RR^{\infty}_{\BF} \leq 1.5$ for the general case: Under adversarial order, much hardness comes from relatively large items of size more than $1/3$; in fact, if all items have size at most $1/3$, an easy argument shows $\nicefrac{4}{3}$-competitiveness even for adversarial arrival order \cite{DBLP:journals/siamcomp/JohnsonDUGG74}.

Moreover, it is natural to ask for the performance in terms of absolute random order ratio.
It is a surprising and rather recent result that for \bestfit, absolute and asymptotic ratios coincide.
The result of \cite{DBLP:conf/soda/Kenyon96} has vast additive terms and it seems that new techniques are required for insights into the absolute random order ratio. In Section~\ref{sec:13largeFinalProof}, we investigate the absolute random order ratio 
for items larger than $1/3$ and obtain the following result.

\begin{proposition}
	\label{prop:largeItemsUBAbs}
	For any list $I$ of items larger than $1/3$, we have 
	$\E[\BF(I^\sigma)] \leq \frac{21}{16} \OPT(I)$.
\end{proposition}
The upper bound of $21/16$ is complemented by the following lower bound.
\begin{proposition}
	\label{prop:largeItemsLB}
	There is a list $I$ of items larger than $1/3$ with $\E[\BF(I^\sigma)] > \frac{6}{5} \OPT(I)$.
\end{proposition}
The proof of Proposition~\ref{prop:largeItemsLB} is given in Appendix~\ref{app:largeItemsLB}.

We also make progress on the hardness side in the general case, which is presented in Section~\ref{sec:lowerBounds}. First, we show that the asymptotic random order ratio is larger than 1.10, improving the previous lower bound of 1.08 from \cite{DBLP:conf/soda/Kenyon96}.
\begin{theorem}
	\label{theo:asymptoticLB}
	The asymptotic random order ratio of \bestfit is $RR_{\BF}^\infty > 1.10$.
\end{theorem}
As it is typically challenging to obtain lower bounds in the random order model, we exploit the connection to the i.i.d.-model. Here, items are drawn independently and identically distributed according to a fixed probability distribution. By defining an appropriate distribution, the problem can be analyzed using Markov chain techniques.
Moreover, we present the first lower bound on the absolute random order ratio:
\begin{theorem}
	\label{theo:absoluteLB}
	The absolute random order ratio of \bestfit is $RR_{\BF} \geq 1.30$.
\end{theorem}
Interestingly, our lower bound on the absolute random order ratio is notably larger than in the asymptotic case (see \cite{DBLP:conf/soda/Kenyon96} and Theorem~\ref{theo:asymptoticLB}).
This suggests either 
\begin{itemize}
	\item a significant discrepancy between $RR_{\BF}$ and $RR_{\BF}^\infty$, which is in contrast to the adversarial setting ($R_{\BF} = R_{\BF}^\infty$, see Table~\ref{tab:relatedWork}), or
	\item a disproof of the conjecture $RR^\infty_{\BF} \approx 1.15$ mentioned in \cite{DBLP:conf/soda/Kenyon96,coffman2013bin}.
\end{itemize}


\section{Notation}
\label{sec:preliminaries}

We consider a list $I=(x_1,\ldots,x_n)$ of $n$ items throughout the paper.
Due to the online setting, $I$ is revealed in \textit{rounds} $1,\ldots,n$.
In round $t$, item $x_t$ arrives and in total, the prefix list $I(t):=(x_1,\ldots,x_t)$ is revealed to the algorithm. The items in $I(t)$ are called the \textit{visible} items of round $t$.
We use the symbol $x_t$ for the item itself and its size $x_t \in (0,1]$ interchangeably. An item $x_t$ is called
\textit{large} (L) if $x_t > 1/2$,
\textit{medium} (M) if $x_t \in \left( 1/3, 1/2 \right]$, and
\textit{small} (S) if $x_t \leq 1/3$.
We also say that $x_t$ is $\alpha$-large if $x_t > \alpha$.

Bins contain items and therefore can be represented as sets. As a bin usually can receive further items in later rounds, the following terms refer always to a fixed round.
We define the \textit{load} of a bin $\mathcal{B}$ as $\sum_{x_i \in \mathcal{B}} x_i$.
Sometimes, we classify bins by their internal structure. We say $\mathcal{B}$ is of \textit{configuration LM} (or $\mathcal{B}$ is an \textit{LM-bin}) if it contains one large and one medium item. The configurations L, MM, etc. are defined analogously. Moreover, we call $\mathcal{B}$ a $k$-bin if it contains exactly $k$ items.
If a bin cannot receive further items in the future, it is called \textit{closed}; otherwise, it is called \textit{open}.

The number of bins which \bestfit uses to pack a list $I$ is denoted by $\BF(I)$. We slightly abuse the notation and refer to the corresponding packing by $\BF(I)$ as well whenever the exact meaning is clear from the context. 
Similarly, we denote by $\OPT(I)$ the number of bins and the corresponding packing of an optimal offline solution.

Finally, for any natural number $n$ we define $[n]:= \{1,\ldots,n\}$. 
Let $\mathcal{S}_n$ be the set of permutations in $[n]$. 
If not stated otherwise, $\sigma$ refers to a permutation drawn uniformly at random from $\mathcal{S}_n$.


\section{Upper bound for \nicefrac{1}{3}-large items}
\label{sec:13largeItems}

In this section, we consider the case where $I$ contains no small items, i.e., where all items are $\nicefrac{1}{3}$-large.
We develop the technical foundations in Sections\nobreakspace  \ref {sec:monotonicity} to\nobreakspace  \ref {sec:13largeGoodOrder}. The final proofs of Theorem\nobreakspace \ref {theo:largeItemsUB} and Proposition\nobreakspace \ref {prop:largeItemsUBAbs} are presented in Section\nobreakspace \ref {sec:13largeFinalProof}.

\subsection{Monotonicity}
\label{sec:monotonicity}

We first define the notion of monotone algorithms.

\begin{definition}
	We call an algorithm \textit{monotone} if increasing the size of one or more items cannot decrease the number of bins used by the algorithm.	
\end{definition}

One might suspect that any reasonable algorithm is monotone. While this property holds for an optimal offline algorithm and some online algorithms as Next Fit \cite{DBLP:books/daglib/0070175}, \bestfit is not monotone in general \cite{DBLP:journals/dam/Murgolo88}.
As a counterexample, consider the lists
\begin{align*}
I &= (0.36, ~ 0.65, ~ \mathbf{0.34}, ~ 0.38, ~ 0.28, ~ 0.35, ~ 0.62)  ~~\text{and} \\
I' &= (0.36, ~ 0.65, ~ \mathbf{0.36}, ~ 0.38, ~ 0.28, ~ 0.35, ~ 0.62) \,.
\end{align*}
Before arrival of the fifth item, $\BF(I(4))$ uses the two bins $\{0.36, 0.38\}$ and $\{0.65, 0.34\}$,
while $\BF(I'(4))$ uses three bins $\{0.36, 0.36\}$, $\{0.65\}$, and $\{0.38\}$.
Now, the last three items fill up the existing bins in $\BF(I'(4))$ exactly. 
In contrast, these items open two further bins in the packing of $\BF(I(4))$. 
Therefore, $\BF(I) = 4 > 3 = \BF(I')$.

However, we can show that \bestfit is monotone for the case of $\nicefrac{1}{3}$-large items. Interestingly, $1/3$ seems to be the threshold for the monotonicity of \bestfit: As shown in the counterexample from the beginning of this section, it is sufficient to have one item $x \in \left( 1/4, 1/3 \right]$ to force \bestfit into anomalous behavior. Anyway, we have the following proposition.

\begin{proposition}
	\label{prop:monotonicity}
	Given a list $I$ of items larger than $1/3$ and a list $I'$ obtained from $I$ by increasing the sizes of one or more items, we have $\BF(I) \leq \BF(I')$.
\end{proposition}

We provide the proof of Proposition\nobreakspace \ref {prop:monotonicity} in Appendix\nobreakspace \ref {app:monotonicity}.
Enabled by the monotonicity, we can reduce an instance of $\nicefrac{1}{3}$-large items to an instance of easier structure. This construction is described in the following.

\subsection{Simplifying the instance}
\label{sec:13largeSimplifying}
Let $I$ be a list of items larger than 1/3. Note that both the optimal and the \bestfit packing use only bins of configurations L, LM, MM, and possibly one M-bin.
However, we can assume a simpler structure without substantial implications on the competitiveness of \bestfit.

\begin{figure}
	\captionsetup[subfigure]{justification=centering}
	\begin{subfigure}[b]{0.53\textwidth}
		\centering
		\includegraphics[]{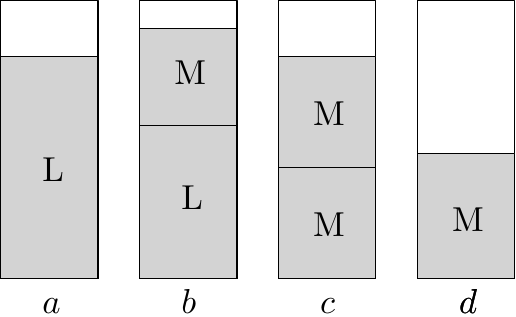}
		\caption{$\OPT(I_0) = \OPT(I_1)$}
		\label{fig:simplifyingI0}
	\end{subfigure}
	\begin{subfigure}[b]{0.25\textwidth}
		\centering
		\includegraphics[]{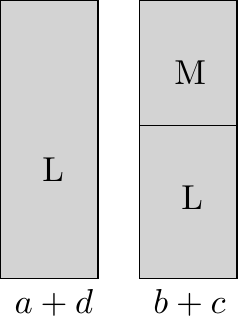}
		\caption{$\OPT(I_2)$}
		\label{fig:simplifyingI2}
	\end{subfigure}
	\begin{subfigure}[b]{0.19\textwidth}
		\centering
		\includegraphics[]{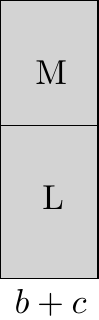}
		\caption{$\OPT(I_3)$}
		\label{fig:simplifyingI3}
	\end{subfigure}
	\caption{Construction from Lemma\nobreakspace \ref {lemma:simplifying} to eliminate L-, MM-, and M-bins in the optimal packing.}
	\label{fig:simplifying}
\end{figure}

\begin{lemma}
	\label{lemma:simplifying}
	Let $I$ be any list that can be packed optimally into $\OPT(I)$ LM-bins.
	If \bestfit has (asymptotic or absolute) approximation ratio $\alpha$ for $I$, then it has (asymptotic or absolute) approximation ratio $\alpha$ for any list of items larger than $1/3$ as well.
\end{lemma}
\begin{proof}
	Let $I_0$ be a list of items larger than $1/3$ and let $a$, $b$, $c$, and $d \leq 1$ be the number of bins in $\OPT(I_0)$ with configurations L, LM, MM, and M, respectively (see Figure\nobreakspace \ref {fig:simplifyingI0}).
	In several steps, we eliminate L-,MM-, and M-bins from $\OPT(I_0)$ while making the instance only harder for \bestfit.
	
	First, we obtain $I_1$ from $I_0$ by replacing items of size $1/2$ by items of size $1/2 - \varepsilon$. 
	By choosing $\varepsilon > 0$ small enough, i.e.,
	$\varepsilon < \min \{ \delta^+ - 1/2, 1/2 - \delta^- \}$, where $\delta^+ = \min \{x_i \mid x_i > 1/2 \}$ and 
	$\delta^- = \max \{x_i \mid x_i < 1/2 \}$, it is ensured that \bestfit packs all items in the same bins than before the modification. Further, the modification does not decrease the number of bins in an optimal packing, so we have
	$\BF(I_0)=\BF(I_1)$ and $\OPT(I_0) = \OPT(I_1)$.
	Now, we obtain $I_2$ from $I_1$ by increasing item sizes: We replace each of the $a+d$ items packed in 1-bins in $\OPT(I_1)$ by large items of size 1. Moreover, any 2-bin (MM or LM) in $\OPT(I_1)$ contains at least one item smaller than $1/2$. These items are enlarged such that they fill their respective bin completely.
	Therefore, $\OPT(I_2)$ has $a+d$ L-bins and $b+c$ LM-bins (see Figure\nobreakspace \ref {fig:simplifyingI2}). We have $\OPT(I_2) = \OPT(I_1)$ and, by Proposition\nobreakspace \ref {prop:monotonicity}, $\BF(I_2) \geq \BF(I_1)$.
	Finally, we obtain $I_3$ from $I_2$ by deleting the $a+d$ items of size 1.
	As size-1 items are packed separately in any feasible packing, $\OPT(I_3) = \OPT(I_2) - (a+d)$ and $\BF(I_3) = \BF(I_2) - (a+d)$.
	Note that $\OPT(I_3)$ contains only LM-bins (see Figure\nobreakspace \ref {fig:simplifyingI3}) and, by assumption, \bestfit has (asymptotic or absolute) approximation ratio $\alpha$ for such lists. Therefore, in general we have a factor $\alpha \geq 1$ and an additive term $\beta$ such that
	$\BF(I_3) \leq \alpha \OPT(I_3) + \beta$. It follows that
	\[
	\BF(I_0) 
	\leq \BF(I_2) 
	= \BF(I_3) + (a+d) 
	\leq \alpha \OPT(I_3) + (a+d) + \beta
	\leq \alpha \OPT(I_0) + \beta \,,
	\]
	which concludes the proof.
\end{proof}
\noindent
By Lemma\nobreakspace \ref {lemma:simplifying}, we can impose the following constraints on $I$ without loss of generality.

\paragraph{Assumption.}
For the remainder of the section, we assume that the optimal packing of $I$ has $k = \OPT(I)$ LM-bins.
For $i \in [k]$, let $l_i$ and $m_i$ denote the large item and the medium item in the $i$-th bin, respectively.
We call $\{l_i, m_i\}$ an \textit{LM-pair}.

\subsection{Good order pairs}
\label{sec:13largeGoodOrder}

If the adversary could control the order of items, he would send all medium items first, followed by all large items. This way, \bestfit opens $k/2$ MM-bins and $k$ L-bins and therefore is 1.5-competitive.
In a random permutation, we can identify structures with a positive impact on the \bestfit packing.
This is formalized in the following random event.
\begin{definition}
	Consider a fixed permutation $\pi \in \mathcal{S}_n$.
	We say that the LM-pair $\{l_i, m_i\}$ arrives in \textit{good order} (or is a \textit{good order pair}) if $l_i$ arrives before $m_i$ in $\pi$.
\end{definition}
Note that in the adversarial setting, no LM-pair arrives in good order, while in a random permutation, this holds for any LM-pair independently with probability $1/2$.
The next lemma is central for the proof of Theorem\nobreakspace \ref {theo:largeItemsUB}. It shows that the number of LM-pairs in good order bound the number of LM-bins in the final \bestfit packing from below.

\begin{lemma}
	\label{lemma:goodOrderPairsLMbins}	
	Let $\pi \in \mathcal{S}_n$ be any permutation and let $X$ be the number of LM-pairs arriving in good order in $I^\pi$.
	The packing $\BF(I^\pi)$ has at least $X$ LM-bins.
\end{lemma}

To prove Lemma\nobreakspace \ref {lemma:goodOrderPairsLMbins}, we model the \bestfit packing by the following bipartite graph:
Let $G_t = (\mathcal{M}_t \cup \mathcal{L}_t, E^{\BF}_t \cup E^{\OPT}_t)$, where $\mathcal{M}_t$ and $\mathcal{L}_t$ are the sets of medium and large items in $I^\pi(t)$, respectively. The sets of edges represent the LM-matchings in the \bestfit packing and in the optimal packing at time $t$, i.e.,
\begin{align*}
E^{\BF}_t &=
\begin{multlined}[t][10.5cm]
\bigl\{ \{m,l \} \in (\mathcal{M}_t \times \mathcal{L}_t) \mid \text{$m$ and $l$ are packed into the same bin} \text{ in $\BF(I^\pi(t))$} \bigr\}
\end{multlined} \\
E^{\OPT}_t &= \bigl\{ \{m_i,l_i \} \in (\mathcal{M}_t \times \mathcal{L}_t) \mid i \in [k] \bigr \} \,.
\end{align*}
We distinguish OPT-edges in good and bad order, according to the corresponding LM-pair.
Note that $G_t$ is not necessarily connected and may contain parallel edges.
We illustrate the graph representation by a small example.

\begin{figure}
	\captionsetup[subfigure]{justification=centering}
	\centering
	\begin{subfigure}[b]{0.5\textwidth}
		\centering
		\includegraphics[]{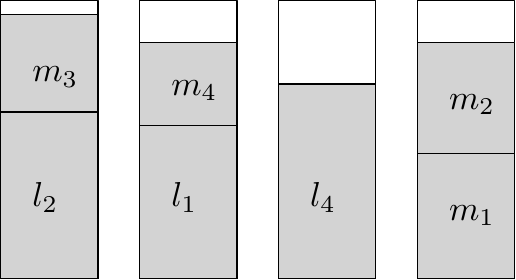}
		\caption{\bestfit packing $\BF(I(7))$}
		\label{fig:goodOrderPacking}
	\end{subfigure}%
	\begin{subfigure}[b]{0.5\textwidth}
		\centering
		\includegraphics[]{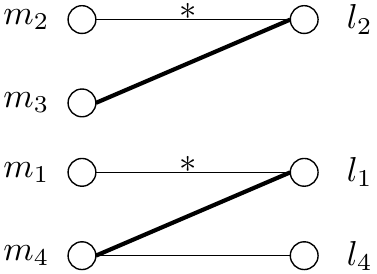}	
		\caption{Graph $G_7$}
		\label{fig:goodOrderGraph}
	\end{subfigure}
	\caption{Visualization of Example\nobreakspace \ref {ex:goodOrderList}. 
		In Figure\nobreakspace \ref {fig:goodOrderGraph}, BF-edges are solid, while OPT-edges are thin. An asterisk indicates an OPT-edge in good order.}
	\label{fig:goodOrderExample}
\end{figure}

\begin{example}
	\label{ex:goodOrderList}
	Let $\eps > 0$ be sufficiently small and define for $i \in [4]$ large items $l_i = 1/2 + i \eps$ and medium items $m_i = 1/2 - i \eps$. Consider the list $I^\pi = (l_2, l_1, m_3, m_4, l_4, m_1, m_2, l_3)$. 
	Figures\nobreakspace \ref {fig:goodOrderPacking} and\nobreakspace  \ref {fig:goodOrderGraph} show the \bestfit packing and the corresponding graph $G_7$ before arrival of the last item. Note that $I^\pi$ has two good order pairs ($\{l_1,m_1\}$ and $\{l_2,m_2\}$) and, according to Lemma\nobreakspace \ref {lemma:goodOrderPairsLMbins}, the packing has two LM-bins.
\end{example}

\noindent
The proof of Lemma\nobreakspace \ref {lemma:goodOrderPairsLMbins} essentially boils down to the following claim:

\begin{claim}
	\label{claim:connectedComponentsGraph}
	In each round $t$ and in each connected component $C$ of $G_t$, the number of BF-edges in $C$ is at least the number of OPT-edges in good order in $C$.
\end{claim} 
We first show how Lemma\nobreakspace \ref {lemma:goodOrderPairsLMbins} follows from \MakeUppercase Claim\nobreakspace \ref {claim:connectedComponentsGraph}. Then, we work towards the proof of \MakeUppercase Claim\nobreakspace \ref {claim:connectedComponentsGraph}.

\begin{proof}[Proof of Lemma\nobreakspace \ref {lemma:goodOrderPairsLMbins}]
		\MakeUppercase Claim\nobreakspace \ref {claim:connectedComponentsGraph} implies that in $G_n$, the total number of BF-edges (summed over all connected components) is at least $X$. Therefore, the packing has at least $X$ LM-bins and thus not less than the number of good order pairs $X$.
\end{proof}

\noindent
Before proving \MakeUppercase Claim\nobreakspace \ref {claim:connectedComponentsGraph}, we show the following property of $G_t$.

\begin{claim}
	\label{claim:alternatingPathOrder}
	Consider the graph $G_t$ for some $t \in [n]$.
	Let $Q=(b_w,a_{w-1},b_{w-1}, \allowbreak,\ldots,a_1,b_1)$ with $w \geq 1$ be a maximal alternating path such that
	$\{a_j, b_j\}$ is an OPT-edge in good order and $\{a_j, b_{j+1} \}$ is a BF-edge for any $j \in [w-1]$
	(i.e., $a$-items and $b$-items represent medium and large items, respectively).
	It holds that $b_w \geq b_1$.
\end{claim}

\begin{proof}
	
	We show the claim by induction on $w$.
	Note that the items' indices only reflect the position along the path, not the arrival order.
	For $w=1$, we have $Q=(b_w)=(b_1)$ and thus, the claim holds trivially.
	
	Now, fix $w \geq 2$ and suppose that the claim holds for all paths $Q'$ with $w' \leq w-1$.
	We next prove $b_w \geq b_1$.
	Let $t' \leq t$ be the arrival time of the $a$-item $a_d$ that arrived latest among all $a$-items in $Q$. We consider the graph $G_{t'-1}$, i.e., the graph immediately before arrival of $a_d$ and its incident edges.
	Note that in $G_{t'-1}$, all items $a_i$ with $i \in [w-1] \setminus \{d\}$ and $b_i$ with $i \in [w-1]$ are visible.
	Let $Q'=(b_w,\ldots,a_{d+1},b_{d+1})$ and $Q''=(b_d,\ldots, a_1, b_1)$ be the connected components of $b_w$ and $b_1$ in $G_{t'-1}$.
	As $Q'$ and $Q''$ are maximal alternating paths shorter than $Q$, we obtain from the induction 
	hypothesis $b_w \geq b_{d+1}$ and $b_d \geq b_1$. 
	Note that $b_{d+1}$ and $b_1$ were visible and packed into L-bins on arrival of $a_d$.
	Further, $a_d$ and $b_1$ would fit together, as $a_d + b_1 \leq a_d + b_d \leq 1$. However, \bestfit packed $a_d$ with $b_{d+1}$, implying $b_{d+1} \geq b_1$.
	Combining the inequalities yields $b_w \geq b_{d+1} \geq b_1$, concluding the proof.
\end{proof}

\noindent
Now, we are able to prove the remaining technical claim.

\begin{proof}[Proof of \MakeUppercase Claim\nobreakspace \ref {claim:connectedComponentsGraph}]
		Note that the number of OPT-edges in good order can only increase on arrival of a medium item $m_i$ where $\{m_i, l_i\}$ is an LM-pair in good order.
		Therefore, it is sufficient to verify \MakeUppercase Claim\nobreakspace \ref {claim:connectedComponentsGraph} in rounds $t_1 < \ldots < t_j$ such that in round $t_i$, item $m_i$ arrives and $l_i$ arrived previously.
		
		\paragraph{Induction base.}
		In round $t_1,$ there is one OPT-edge $\{m_1,l_1\}$ in good order. We need to show that there exists at least one BF-edge in $G_{t_1}$, or, alternatively, at least one LM-bin in the packing. 
		If the bin of $l_1$ contains a medium item different from $m_1$, we identified one LM-bin. 
		Otherwise, \bestfit packs $m_1$ together with $l_1$ or some other large item, again creating an LM-bin.
		
		\paragraph{Induction hypothesis.}
		Fix $i \geq 2$ and assume that \MakeUppercase Claim\nobreakspace \ref {claim:connectedComponentsGraph} holds up to round $t_{i-1}$. 
		
		\paragraph{Induction step.}
		We only consider the connected component of $m_i$, as by the induction hypothesis, the claim holds for all remaining connected components. If $m_i$ is packed into an LM-bin, the number of BF-edges increases by one and the claim holds for round $t_i$. Therefore, assume that $m_i$ is packed by \bestfit in an M- or MM-bin.
		This means that in $G_{t_i}$, vertex $m_i$ is incident to an OPT-edge in good order, but not incident to any BF-edge.
		Let $P=(m_i, l_i, \ldots,v)$ be the maximal path starting from $m_i$ alternating between OPT-edges and BF-edges.
		
		\paragraph{Case 1: $v$ is a medium item.}
		For illustration, consider Figure\nobreakspace \ref {fig:goodOrderGraph} with $m_i = m_2$ and $v = m_3$.
		Since $P$ begins with an OPT-edge and ends with a BF-edge, the number of BF-edges in $P$ equals the number of OPT-edges in $P$. The latter number is clearly at least the number of OPT-edges in good order in $P$.
		
		\paragraph{Case 2: $v$ is a large item.}
		For illustration, consider Figure\nobreakspace \ref {fig:goodOrderGraph} with $m_i = m_1$ and $v = l_4$.
		We consider two cases. If $P$ contains at least one OPT-edge which is not in good order, the claim follows for the same argument as in Case 1.
		
		Now, suppose that all OPT-edges in $P$ are in good order.
		Let $P'$ be the path obtained from $P$ by removing the item $m_i$.
		As $P'$ satisfies the premises of \MakeUppercase Claim\nobreakspace \ref {claim:alternatingPathOrder}, we obtain $l_i \geq v$.
		This implies that $m_i$ and $v$ would fit together, as $m_i + v \leq m_i + l_i \leq 1$.
		However, $m_i$ is packed in an M- or MM-bin by assumption, although $v$ is a feasible option on arrival of $m_i$.
		As this contradicts the \bestfit rule, we conclude that case 2 cannot happen.
\end{proof}

\subsection{Final proofs}
\label{sec:13largeFinalProof}

Finally, we prove the main result of this section.
	\begin{proof}[Proof of Theorem\nobreakspace \ref {theo:largeItemsUB}]
		Let $X$ be the number of good order pairs in $I^\sigma$ and let $Y$ be the number of LM-bins in the packing $\BF(I^\sigma)$. We have $Y \geq X$ by Lemma\nobreakspace \ref {lemma:goodOrderPairsLMbins}. For the remaining large and medium items, \bestfit uses $(k-Y)$ L-bins and $\lceil (k-Y)/2 \rceil$ MM-bins (including possibly one M-bin), respectively.
		Therefore, 
		\begin{equation}
		\label{eq:largeItemsXLMbins}
		\BF(I^\sigma) 
		= Y + (k-Y) + \left\lceil \frac{k-Y}{2} \right \rceil 
		\leq k + \left\lceil \frac{k-X}{2} \right \rceil
		= \frac{3k}{2} - \frac{X}{2} + \frac{\xi(X)}{2} \,,
		\end{equation}
		where $\xi(X) = (k-X) \bmod 2$.
		Using linearity and monotonicity of expectation, we obtain
		\begin{equation}
		\label{eq:largeItemsXLMbins2}
		\E[\BF(I^\sigma)]
		\leq \frac{3k}{2} - \frac{\E[X]}{2} + \frac{\Pr [\xi(X)=1]}{2} \,.
		\end{equation}
		Since $\sigma$ is uniformly distributed on $\mathcal{S}_n$, each LM-pair arrives in good order with probability $1/2$. Therefore, $\E[X]= k/2$ and $\Pr[\xi(X)=1] = 1/2$. Hence, 
		\begin{equation}
		\label{eq:largeItems54+Constant}
		\E[\BF(I^\sigma)] \leq \frac{3k}{2} - \frac{k/2}{2} + \frac{1/2}{2} = \frac{5k}{4} + \frac{1}{4} 
		= \frac{5}{4} \OPT(I) + \frac{1}{4}
		\,,
		\end{equation}
		where we used $k = \OPT(I)$. This concludes the proof.
\end{proof}
To obtain the upper bound of $21/16$ on the absolute random order ratio (Proposition\nobreakspace \ref {prop:largeItemsUBAbs}), we analyze a few special cases more carefully.

\begin{proof}[Proof of Proposition\nobreakspace \ref {prop:largeItemsUBAbs}]
		For $k \geq 4$, the claim follows immediately from Equation\nobreakspace \textup {(\ref {eq:largeItems54+Constant})}:
		\[
		\frac{\E[\BF(I^\pi)]}{\OPT(I)} = \frac{(5k)/4 + 1/4}{k} = \frac{5}{4} + \frac{1}{4k} \leq \frac{21}{16} \,.
		\]
		Since \bestfit is clearly optimal for $k=1$, it remains to verify the cases $k \in \{2,3\}$.
		\begin{description}
			\item[$k=2$] It is easily verified that there are 16 out of $4!=24$ permutations where \bestfit is optimal and that it opens at most 3 bins otherwise. 
			Therefore,
			\[
			\E[\BF(I^\sigma)] = \frac{1}{4!} \cdot \left( 16 \OPT(I) + 8 \cdot \frac{3}{2} \OPT(I) \right) = \frac{7}{6} \OPT(I) < \frac{21}{16} \OPT(I).
			\] 
			\item[$k=3$] When $k$ is odd, there must be at least one LM-bin in the \bestfit packing: Suppose for contradiction that all M-items are packed in MM- or M-bins. As $k$ is odd, there must be an item $m_i$ packed in an M-bin.
			If $l_i$ arrives before $m_i$, item $l_i$ is packed in an L-bin, as there is no LM-bin.
			Therefore, \bestfit packs $m_i$ with $l_i$ or some other L-item instead of opening a new bin. 
			If $l_i$ arrives after $m_i$, \bestfit packs $l_i$ with $m_i$ or some other M-item. 
			We have a contradiction in both cases.
			
			Therefore, for $k=3$ we have at least one LM-bin, even if no LM-pair arrives in good order.
			Consider the proof of Theorem\nobreakspace \ref {theo:largeItemsUB}. Instead of $Y \geq X$, we can use the stronger bound $Y \geq X'$ with $X' := \max \{1, X\}$ on the number of LM-bins. The new random variable satisfies $\E[X'] = k/2 + 1/2^k$ and $\Pr[\xi(X') = 1] = 1/2 - 1/2^k$. Adapting Equations\nobreakspace \textup {(\ref {eq:largeItemsXLMbins})} and\nobreakspace  \textup {(\ref {eq:largeItemsXLMbins2})} appropriately, we obtain
			\begin{align*}
			\frac{\E[\BF(I^\sigma)]}{\OPT(I)}
			&= \frac{1}{k} \cdot \left( \frac{3k}{2} - \frac{k/2 + 1/2^k}{2} + \frac{1/2 - 1/2^k}{2} \right) \\
			&= \frac{5}{4} + \frac{1}{4k} - \frac{1}{k 2^k}
			= \frac{31}{24} 
			< \frac{21}{16} \,.
			\end{align*}
		\end{description}
\end{proof}


\section{Lower bounds}
\label{sec:lowerBounds}

In this section, we present the improved lower bound on $RR_{\BF}^\infty$ (Theorem\nobreakspace \ref {theo:asymptoticLB}) and the first lower bound on the absolute random order ratio $RR_{\BF}$ (Theorem\nobreakspace \ref {theo:absoluteLB}).

\subsection{Asymptotic random order ratio}

Another model of probabilistic analysis is the i.i.d.-model, where the input of the algorithm is a sequence of independent and identically distributed (i.i.d.) random variables. 
Here, the performance measure of algorithm $\A$ is 
${\E[\A(I_n(F))]} / {\E[\OPT(I_n(F))]}$, where $I_n(F):=(X_1,\ldots,X_n)$ is a list of $n$ random variables drawn i.i.d.\ according to $F$.
This model is in general weaker than the random order model, which is why lower bounds in the random order model can be obtained from the i.i.d.-model. This is formalized in the following lemma.

\begin{lemma}
	\label{lemma:IIDvsROM}	
	Consider any online bin packing algorithm $\mathcal{A}$.
	Let $F$ be a discrete distribution and $I_n(F) = (X_1,\ldots,X_n)$ be a list of i.i.d.\ samples.
	For $n \to \infty$, there exists a list $I$ of $n$ items such that
	\[
	\frac{\E[\A(I^\sigma)]}{\OPT(I)} \geq \frac{\E[\A(I_n(F))]}{\E[\OPT(I_n(F))]} \,.
	\]
	Moreover, if there exists a constant $c > 0$ such that $X_i \geq c$ for all $i \in [n]$, we have $\OPT(I) \geq cn$.
\end{lemma}
This technique has already been used in \cite{DBLP:conf/soda/Kenyon96} to establish the lower bound of 1.08, however, without a formal proof.
Apparently, the only published proofs of this connection address bin covering \cite{DBLP:conf/latin/FischerR16,DBLP:journals/tcs/ChristFL14}. 
We provide a constructive proof of Lemma\nobreakspace \ref {lemma:IIDvsROM} in Appendix\nobreakspace \ref {app:asymptoticLB} for completeness.
The improved lower bound from Theorem\nobreakspace \ref {theo:asymptoticLB} now follows by combining Lemma\nobreakspace \ref {lemma:IIDvsROM} with the next lemma.

\begin{figure}[t]
	\captionsetup[subfigure]{justification=centering}
	\begin{subfigure}[b]{0.5\textwidth}
		\resizebox{\textwidth}{!}{%
			\begin{tikzpicture}[shorten >=1pt,auto,node distance=2.4cm]
			
			\node[state] (A) [] {$\mathsf{A}$};
			\node[state] (B) [above right of=A]  {$\mathsf{B}$};
			\node[state] (C) [below right of=A]  {$\mathsf{C}$};
			\node[state] (D) [above right of=B]  {$\mathsf{D}$};
			\node[state] (E) [below right of=B]  {$\mathsf{E}$};
			\node[state] (F) [right of=C]  {$\mathsf{F}$};
			\node[state] (G) [right of=D]  {$\mathsf{G}$};
			\node[state] (H) [below of=G]  {$\mathsf{H}$};			
			\node[state] (I) [below of=H]  {$\mathsf{I}$};

			\path[->] 
			(A) edge[very thick] node {$p$} (B)
			(A) edge[very thick] node {$q$} (C)
			(B) edge node {$p$} (D)
			(B) edge node {$q$} (E)
			(C) edge node {$p$} (E)
			(C) edge[below] node {$q$} (F)					
			(D) edge node {$p$} (G)
			(D) edge[above, bend right=40] node {$q$} (A)
			(E) edge[above] node {$1$} (A)
			(F) edge[bend left=80, pos=0.75] node {$1$} (A)
			(G) edge[very thick] node {$q$} (H)
			(G) edge[above, bend right=100] node {$p$} (A)
			(H) edge[bend left=10] node {$p$} (C)
			(H) edge node {$q$} (I)
			(I) edge node {$p$} (F)
			(I) edge[bend right=40] node {$q$} (G);					
			\end{tikzpicture}}
		\caption{Transition diagram}
		\label{fig:markovChainDiagram}
	\end{subfigure}
	\hfill
	\begin{subfigure}[b]{0.33\textwidth}
		\begin{tabular}{llll} \toprule
			State & Load of & \\
			& open bin(s) \\ \midrule
			$\mathsf{A}$   & -- \\ 
			$\mathsf{B}$   & 1/4    \\			
			$\mathsf{C}$   & 1/3    \\
			$\mathsf{D}$  & 2/4    \\
			$\mathsf{E}$   & 7/12    \\			
			$\mathsf{F}$   & 2/3    \\			
			$\mathsf{G}$   & 3/4    \\
			$\mathsf{H}$   & 3/4, 1/3    \\			
			$\mathsf{I}$   & 3/4, 2/3    \\	\bottomrule
		\end{tabular} 	
		\vspace*{20pt}
		\caption{Description of states}
		\label{fig:markovChainStates}
		
	\end{subfigure}
	\caption{Markov chain from Lemma\nobreakspace \ref {lemma:110Distribution}. Bold arcs in Figure\nobreakspace \ref {fig:markovChainDiagram} indicate transitions where \bestfit opens a new bin.}
	\label{fig:markovChain}
\end{figure}

\begin{lemma}
	\label{lemma:110Distribution}
	There exists a discrete distribution $F$ such that for $n \to \infty$, we have
	$\E[\BF(I_n(F))] > \frac{11}{10} \E[\OPT(I_n(F))]$ and each sample $X_i$ satisfies $X_i \geq 1/4$.
\end{lemma}
\begin{proof}
	Let $F$ be the discrete distribution which gives an item of size $1/4$ with probability $p$ and an item of size $1/3$ with probability $q:=1-p$.
	First, we analyze the optimal packing. Let $N_{4}$ and $N_{3}$ be the number of items with size $1/4$ and $1/3$ in $I_n(F)$, respectively. We have
	\[
	\E[\OPT(I_n(F))] 
	\leq \E \left[ \frac{N_{4}}{4} + \frac{N_{3}}{3} + 2 \right]  
	= \frac{np}{4} + \frac{nq}{3} + 2 
	= n \left( \frac{1}{3} - \frac{p}{12} + \frac{2}{n} \right) \,.
	\]
	
	Now, we analyze the expected behavior of \bestfit for $I_n(F)$. 	
	As the only possible item sizes are $1/4$ and $1/3$, we can consider each bin of load more than $3/4$ as closed.
	Moreover, the number of possible loads for open bins is small and \bestfit maintains at most two open bins at any time.
	Therefore, we can model the \bestfit packing by a Markov chain as follows.
	Let the nine states $\mathsf{A},\mathsf{B},\ldots,\mathsf{I}$ be defined as in Figure\nobreakspace \ref {fig:markovChainStates}.
	The corresponding transition diagram is depicted in Figure\nobreakspace \ref {fig:markovChainDiagram}.
	This Markov chain converges to the stationary distribution
	\begin{align*}
	\omega &= 
	(\omega_\mathsf{A},\ldots,\omega_\mathsf{I}) \\
	&= 
	\frac{1}{\lambda}
	\left( 1, ~ p, ~ q+pq\vartheta, ~ p^2, ~ 2pq+p^2q\vartheta, ~ q^2+2pq^2\vartheta, ~ \vartheta, ~ q\vartheta, ~ q^2 \vartheta \right) \,,
	\end{align*}
	where we defined 
	$\vartheta :=\frac{p^3}{1-q^3}$ and  
	$\lambda := \vartheta q \left( 3 - q^2 \right) + \vartheta + 3$. 
	A formal proof of this fact can be found in Appendix\nobreakspace \ref {app:stationaryDistribution}.
	
	Let $V_\mathsf{S}(t)$ denote the number of visits to state $\mathsf{S} \in \{ \mathsf{A}, \ldots, \mathsf{I} \}$ up to time $t$.
	By a basic result from the theory of ergodic Markov chains (see \cite[Sec.\ 4.7]{levin2017markov}), it holds that
	$\lim_{t \to \infty} \frac{1}{t} \cdot V_\mathsf{S}(t) = \omega_S$.
	In other words, the proportion of time spent in state $\mathsf{S}$ approaches its probability $\omega_\mathsf{S}$ in the stationary distribution.
	This fact can be used to bound the total number of opened bins over time.
	Note that \bestfit opens a new bin on the transitions $A \to B$, $A \to C$, and $G \to H$ (see Figure\nobreakspace \ref {fig:markovChainDiagram}). 
	Hence, $\E[\BF(I_n(F))] = V_\mathsf{A}(n) + q V_\mathsf{G}(n)$.
	Setting $p = 0.60$, we obtain finally
	\begin{equation}
	\lim_{n \to \infty} \frac{\E[\BF(I_n(F))]}{\E[\OPT(I_n(F))]}  
	\geq \lim_{n \to \infty} \frac{n \omega_\mathsf{A} + nq \omega_\mathsf{G}}{n \left(\frac{1}{3} - \frac{p}{12} + \frac{2}{n} \right)} 
	= \frac{1 + q \vartheta}{\lambda \cdot \left( \frac{1}{3} - \frac{p}{12} \right)}
	> \frac{11}{10} \,.
	\end{equation}
\end{proof}

\subsection{Absolute random order ratio}
\label{sec:absLB}
Theorem\nobreakspace \ref {theo:absoluteLB} follows from the following lemma.
\begin{lemma}
	There exists a list $I$ such that $\E[\BF(I^\sigma)] = \frac{13}{10} \OPT(I)$.
\end{lemma}

\begin{proof}
	Let $\eps > 0$ be sufficiently small and let $I := (a_1,a_2,b_1,b_2,c)$ where
	\[
	a_1 = a_2 = \frac{1}{3} + 4 \eps, ~~~
	b_1 = b_2 = \frac{1}{3} + 16 \eps, ~~~
	c = \frac{1}{3} - 8 \eps \,.
	\]
	An optimal packing of $I$ has two bins $\{a_1,a_2,c\}$ and $\{b_1,b_2\}$, thus $\OPT(I)=2$.
	Subsequently, we argue that Best Fit needs two or three bins depending on the order of arrival.
	
	Let $E$ be the event that exactly one $b$-item arrives within the first two rounds. 
	After the second item, the first bin is closed, as its load is at least $\frac{1}{3} + 16\eps + \frac{1}{3} - 8\eps = \frac{2}{3} + 8 \eps$.
	Among the remaining three items, there is a $b$-item of size $\frac{1}{3} + 16 \eps$ and at least one $a$-item of size $\frac{1}{3} + 4\eps$. This implies that a third bin needs to be opened for the last item.
	As there are exactly $2 \cdot 3 \cdot 2! \cdot 3! = 72$ permutations where $E$ happens, we have
	$\Pr[E] = \frac{72}{5!} = \frac{3}{5}$.
	
	On the other side, \bestfit needs only two bins if one of the events $F$ and $G$, defined in the following, happen. 
	Let $F$ be the event that both $b$-items arrive in the first two rounds. Then, the remaining three items fit into one additional bin.
	Moreover, let $G$ be the event that the set of the first two items is a subset of $\{a_1,a_2,c\}$. Then, the first bin has load at least $\frac{2}{3} - 4 \eps$, thus no $b$-item can be packed there. Again, this ensures a packing into two bins.
	By counting permutations, we obtain $\Pr[F] = \frac{2! \cdot 3!}{5!} = \frac{1}{10}$ and $\Pr[G] = \frac{3 \cdot 2! \cdot 3!}{5!} = \frac{3}{10}$. 
	
	As the events $E$, $F$, and $G$ partition the probability space, we obtain
	\[
	\frac{\E[\BF(I^\sigma)]}{\OPT(I)} 
	= \frac{\Pr[E] \cdot 3 + \left( \Pr[F] + \Pr[G] \right) \cdot 2}{2} 
	= \frac{\frac{3}{5} \cdot 3 + \left( \frac{1}{10} + \frac{3}{10} \right) \cdot 2}{2} 
	= \frac{13}{10} \,.
	\]
\end{proof}
The construction from the above proof is used in \cite{DBLP:journals/siamcomp/JohnsonDUGG74} to prove that \bestfit is 1.5-competitive under adversarial arrival order if all item sizes are close to $1/3$. Interestingly, it gives a strong lower bound on the absolute random order ratio as well.


\bibliography{literature}
\newpage

\appendix

\section{Monotonicity}
\label{app:monotonicity}

Proposition\nobreakspace \ref {prop:monotonicity} follows by applying the following lemma iteratively.
A technically similar proof appeared in \cite{DBLP:journals/combinatorica/Shor86}, where Shor showed that the MBF algorithm from \cite{DBLP:journals/combinatorica/Shor86} is monotone under removal of items.

\begin{figure}
	\captionsetup[subfigure]{justification=centering}	
	\centering
	\begin{subfigure}{.43\textwidth}
		\centering
		\includegraphics[height=6cm]{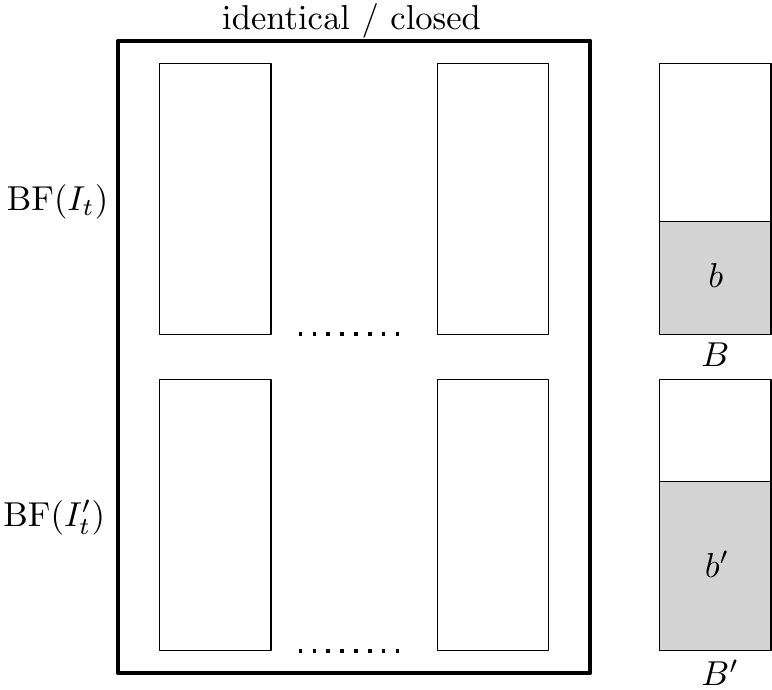}
		\caption{Property $(\ast2)$}
	\end{subfigure}%
	\hfill
	\begin{subfigure}{.57\textwidth}
		\centering
		\includegraphics[height=5.6cm]{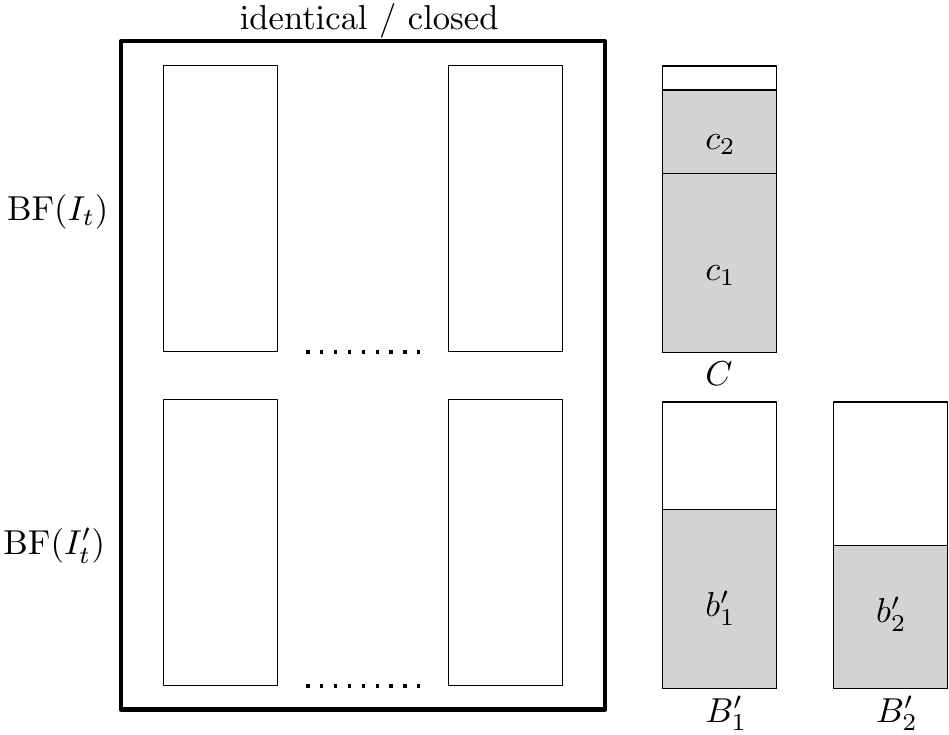}
		\caption{Property $(\ast3)$}
	\end{subfigure}
	\caption{Relating the packings $\BF(I_t)$ and $\BF(I'_t)$ in the proof of Lemma\nobreakspace \ref {lemma:monotonicityShrink}.}
	\label{fig:monotonicity}
\end{figure}

\begin{lemma}
	\label{lemma:monotonicityShrink}
	Let $I=(x_1,\ldots,x_n)$ be any list of items larger than $1/3$.
	Let $I' = (x'_1,\ldots,x'_n)$ with $x'_i > x_i$ for a single $i \in [n]$ and $x'_j = x_j$ for all $j \neq i$.
	We have $\BF(I) \leq \BF(I')$.
\end{lemma}
\begin{proof}
	All bins in any packing of $I$ or $I'$ contain at most two items.
	We call two 1-bins of $\BF(I)$ and $\BF(I')$ \textit{pairwise-identical} if they contain items of the same size.
	Moreover, we call any two 2-bins of $\BF(I)$ and $\BF(I')$ \textit{pairwise-closed}, as neither of the two bins can receive a further item.
	
	For ease of notation, let $I_t = I(t)$ and $I'_t = I'(t)$. We show that at any time $t$, the packings $\BF(I_t)$ and $\BF(I'_t)$ are related in one of three ways (see Figure\nobreakspace \ref {fig:monotonicity}).
	\begin{description}
		\item[$(\ast1)$] All bins are pairwise-identical or pairwise-closed.
		\item[$(\ast2)$] All bins are pairwise-identical or pairwise-closed, except for two 1-bins $B=\{b\}$ and $B'=\{b'\}$ in $\BF(I_t)$ and $\BF(I'_t)$, respectively, where $b < b'$.
		\item[$(\ast3)$] All bins are pairwise-identical or pairwise-closed, except for a 2-bin $C = \{c_1, c_2\}$ in $\BF(I_t)$ which does not exist in $\BF(I'_t)$, and two 1-bins $B'_1 = \{b'_1\}$, $B'_2=\{b'_2\}$ in $\BF(I'_t)$ which do not exist in $\BF(I_t)$.
	\end{description}
	
	Note in all three cases, $\BF(I_t) \leq \BF(I'_t)$.  As this property is maintained until $t=n$, it implies the lemma. Subsequently, we show the claim inductively for each round.
	
	Before round $i$, both lists contain items of identical size and thus, the packing are clearly related by $(\ast1)$.
	In round $i$, both packings deviate possibly, since $x_i < x'_i$. Here, three cases can occur.
	\begin{enumerate}
		\item Both items go into new bins. Then, $(\ast2)$ holds.
		\item Both items go into identical 1-bins. Then, $(\ast1)$ holds.
		
		\item Best-fit packs $x_i$ into an existing 1-bin $B_1$ in $\BF(I_i)$, while it opens a new bin $B'_2$ for $x'_i$ in $\BF(I'_i)$.
		Before packing $x_i$ into $B_1$, there was a 1-bin $B'_1$ in $\BF(I'_{i})$ such that $B_1$ and $B'_1$ were pairwise identical.
		By packing $x_i$, bin $B_1$ becomes a 2-bin. Therefore, $(\ast3)$ holds with $C = B_1$ as 2-bin and $B'_1$, $B'_2$ as 1-bins.
	\end{enumerate}
	Note that these three cases are exhaustive: The algorithm packs either both items into new bins (a), both items into existing 1-bins (b), or $x_i$ into an existing 1-bin and $x'_i$ into a new bin (c). 
	In case (b), the 1-bins must be pairwise-identical as this holds for all bins up to time $i$.
	The inverse situation of (c) cannot occur since $x_i < x'_i$.
	
	For the induction step, we consider round $t \geq i+1$ and suppose that $\BF(I_{t-1})$ and $\BF(I'_{t-1})$ are related either by $(\ast1)$, $(\ast2)$, or $(\ast3)$. Note that in round $t$, in both lists the current item has equal size $x_t = x'_t$ again.
	
	\paragraph{\small Case 1} If $(\ast1)$ holds at time $t-1$, then $(\ast1)$ is maintained in round $t$.
	\paragraph{\small Case 2} Suppose $(\ast2)$ holds at time $t-1$.
	If $x'_t$ is packed in $\BF(I'_t)$ into an existing 1-bin, we know that \bestfit does not open a new bin for $x_t$ in $\BF(I_t)$ as well.
	Depending on the chosen bins, either $(\ast2)$ is maintained, or $(\ast1)$ holds.
	
	Now, assume that $x'_t$ is packed into a new bin in $\BF(I'_t)$. Either, a new bin is opened in $\BF(I_t)$ as well, 
	or $\BF(I_t)$ packs $x_t$ into bin $B$.
	In the first case, $(\ast2)$ is maintained, in the second case $(\ast3)$ holds.
	Note that $\BF(I_t)$ cannot pack $x_t$ into any existing 1-bin other than $B$, as otherwise $\BF(I'_t)$ would pack $x'_t$ into the corresponding identical bin instead of opening a new bin.
	
	\paragraph{\small Case 3} Suppose $(\ast3)$ holds at time $t-1$.
	If \bestfit packs $x_t$ into an existing 1-bin, we know that it packs $x'_t$ into an existing 1-bin as well.
	Hence, in both packings a 1-bin becomes a 2-bin and $(\ast3)$ is maintained.
	
	If \bestfit packs $x_t$ into a new bin, we know that none of the existing 1-bins in $\BF(I'_t)$, except for possibly $B'_1$ or $B'_2$, are suitable for $x'_t$.
	Therefore, \bestfit either opens a new bin to pack $x'_t$ as well (then, $(\ast3)$ is maintained), or it packs $x'_t$ into $B' \in \{B'_1, B'_2\}$.
	Suppose that $B'=B'_1$. Then, all bins in the packings are pairwise-identical or closed, except for $B'_2 = \{b'_2\}$ in $\BF(I'_t)$ and the new bin $\{x_t\}$ in $\BF(I_t)$. Hence, $(\ast2) $ holds if $b'_2 > x_t$. 
	To see this, observe that $b'_1 + b'_2 > 1$ since $b'_1$ and $b'_2$ have been packed into separate bins previously.
	Moreover, since \bestfit packed $x'_t$ with $b'_1$, we have $x_t + b'_1 \leq 1$.
	Therefore, $b'_2 > 1 - b'_1 \geq 1 - (1 - x_t) = x_t$.
	The case $B'=B'_2$ is analogous.
\end{proof}

\section{Lower bound for $\nicefrac{1}{3}$-large items}
\label{app:largeItemsLB}
Here, we prove the existence of a list $I$ with $\nicefrac{1}{3}$-large items and $\E[\BF(I^\sigma)] > \frac{6}{5} \OPT(I)$.

\begin{proof}[Proof of Proposition\nobreakspace \ref {prop:largeItemsLB}]
		We construct a list of $k=3$ LM-pairs. For sufficiently small $\eps>0$ and $i \in [k]$ define
		$l_i = \frac{1}{2} + i \eps$ and $m_i = \frac{1}{2} - i \eps$. This way, $l_1 < l_2 < l_3$ and $m_1 > m_2 > m_3$.
		Clearly, $\OPT(I)=3$. We can show that \bestfit uses 4 instead of 3 bins in at least $440$ permutations.
		Therefore,
		\[
		\frac{\E[\BF(I^\sigma)]}{\OPT(I)} \geq \frac{\frac{1}{6!} \cdot \left( 440 \cdot 4 + (6!-440) \cdot 3 \right)}{3} = \frac{65}{54} > \frac{6}{5} \,.
		\]
		We call the event that \bestfit packs two items different from $\{l_i, m_i\}$ into the same bin a \textit{non-optimal match}.
		This event occurs if \bestfit packs either $l_i$ and $m_j$ for $1 \leq i < j \leq 3$, or two medium items $m_i$ and $m_j$ together into the same bin.
		It is easy to see that any packing with a non-optimal match needs at least 4 bins. However, \bestfit never uses more than 4 bins.
		
		In the following, we partition the set of all permutations with a non-optimal match according to the first time where this happens. Note that the first possibility is after arrival of the second item. Moreover, if no non-optimal match happened before arrival of the fifth item, the fifth and sixth item build an LM-pair and therefore, no non-optimal match occurs at all.
		Let $a_1, \ldots, a_4$ denote the first four items in a fixed permutation.
		\begin{description}
			
			\item[\textit{Case 1: First non-optimal match after the  second item}] ~\\
			We have either $a_1, a_2 \in \{l_i, m_j \}$ with $i < j$, or $a_1, a_2 \in \{m_i, m_j\}$ with $i \neq j$.
			In both cases, we can choose the set $\{a_1, a_2\}$ in three ways, can arrange $a_1$ and $a_2$ in $2!$ ways and the remaining items in $4!$ ways. The total number of such permutations is
			\[
			2 \cdot 3 \cdot 2! \cdot 4! = 288 \,. \]

			\item[\textit{Case 2: First non-optimal match after the third item}] ~\\
			Before $a_3$ arrives and is matched non-optimally, $a_1$ and $a_2$ need to be packed into separate bins and $a_3$ is matched non-optimally.
			This happens in the following cases.
			\begin{description}
				\item[\textit{Case 2a}] $a_1, a_2 \in \{ l_1, l_2 \}$ and $a_3 = m_3$.
				\item[\textit{Case 2b}] $a_1, a_2 \in \{l_1, l_3 \}$ and $a_3 = m_2$.  
				\item[\textit{Case 2c}] $a_1, a_2 \in \{l_2, m_1\}$ and $a_3 = m_3$.
				\item[\textit{Case 2d}] $a_1, a_2 \in \{l_3, m_1\}$ and $a_3 = m_2$.
				\item[\textit{Case 2e}] $a_1, a_2 \in \{m_2, l_3 \}$ and $a_3 \in \{m_1, l_1\}$.  
			\end{description}
			Note that for each case 2a-2d, we have $2! \cdot 3!$ permutations, while in case 2e, in addition, we can choose $a_3$ in one of two ways. Therefore, we have a total number of permutations for case 2 of 
			\[
			4 \cdot (2! \cdot 3!) + 2! \cdot 2 \cdot 3! = 48 + 24 = 72 \,.
			\]
			
			\item[\textit{Case 3: First non-optimal match after the fourth item}]~\\
			Here, we must have an optimal match either after the second item (Case 3a) or after the third item (Case 3b).
			
			\begin{description}
				\item[\textit{Case 3a}] We have $a_1, a_2 \in \{l_i, m_i\}$ for some $i \in [3]$.
				Now, $\{a_3,a_4\}$ can contain either the remaining two medium items, or the larger of the remaining two large items and the smaller of the remaining two medium items. The fifth and sixth item can be arranged in $2!$ ways. 
				The total number of such permutations is 
				\[
				3 \cdot 2! \cdot 2 \cdot 2! \cdot 2! = 48 \,.
				\]
				
				\item[\textit{Case 3b}] We have an optimal match after the third item followed by a non-optimal match after the fourth item. This happens in the following cases.
				\begin{itemize}
					\item $a_1, a_2 \in \{ l_3, m_1 \}$, $a_3 = m_3$,                    $a_4 = m_2$.
					\item $a_1, a_2 \in \{ l_1, l_3 \}$,   $a_3 = m_3$,                     $a_4 = m_2$.			
					\item $a_1, a_2 \in \{ l_2, m_1 \}$, $a_3 \in \{m_2, l_1 \}$,    $a_4 = m_3$.
					\item $a_1, a_2 \in \{ l_1, l_2 \}$,   $a_3 \in \{m_1, m_2 \}$,  $a_4 = m_3$.
					\item $a_1, a_2 \in \{ l_3, m_2 \}$, $a_3 = m_3$,                     $a_4 = \{m_1, l_1\}$.	
				\end{itemize}
				In each of the first two cases, we get $2! \cdot 2!$ permutations.
				In each of the remaining three cases, we have $2! \cdot 2 \cdot 2!$ permutations, since we can choose one additional item among two elements.
				The total number of permutations in case 3b is thus
				\[
				2 \cdot (2! \cdot 2!) + 3 \cdot (2! \cdot 2 \cdot 2!) = 8 + 24  = 32 \,.
				\]
			\end{description}
		\end{description}
\end{proof}

\section{Lower bound on the asymptotic random order ratio}
\label{app:asymptoticLB}
\subsection{Proof of Lemma~\ref{lemma:IIDvsROM}}

Let $\mathcal{I} = \{ I \mid \Pr[I_n(F)=I] > 0 \}$ be the set of possible outcomes of $I_n(F)$.
We say that two lists $I_1, I_2 \in \mathcal{I}$ are \textit{similar} ($I_1 \sim I_2$) if there is a permutation $\pi \in \mathcal{S}_n$ such that $I_1 = I_2^\pi$. Note that $\sim$ defines an equivalence relation on $\mathcal{I}$.
Let $\mathcal{H}$ be a complete set of representatives of $\sim$. This way, $\mathcal{I} = \biguplus_{H \in \mathcal{H}} \{ H^\pi \mid \pi \in \mathcal{S}_n \}$.

We will use the following two technical claims which we will prove later.
\begin{claim}
	\label{claim:CommonProbInEquivClass}
	Let $\ALG \in \{\A, \OPT \}$. For each $H \in \mathcal{H}$, there exists $\lambda_H > 0$ such that
	\begin{equation*}
	\E[\ALG(I_n(F))] = \sum_{H \in \mathcal{H}} \left(\lambda_H \cdot \sum_{\pi \in \mathcal{S}_n} \ALG(H^\pi)\right) \,.
	\end{equation*}
\end{claim}

\begin{claim}
	\label{claim:RatioOfSumsUB}
	For any two nonnegative sequences $(a_i)_{i \in [m]}$ and $(b_i)_{i \in [m]}$, we have 
	\begin{equation*}
	\frac{\sum_{j=1}^{m} a_j}{\sum_{j=1}^{m} b_j} \leq \max_{j \in [m]} \frac{a_j}{b_j} \,.
	\end{equation*}
\end{claim}

\begin{proof}[Proof of Lemma~\ref{lemma:IIDvsROM}]
		Using \MakeUppercase Claims\nobreakspace \ref {claim:CommonProbInEquivClass} and\nobreakspace  \ref {claim:RatioOfSumsUB}, Lemma~\ref{lemma:IIDvsROM} follows from the following reasoning.
		Let 
		\[I = \arg \max_{H \in \mathcal{H}} \frac{\lambda_H \cdot \sum_{\pi \in \mathcal{S}_n} \A(H^\pi)}{\lambda_H \cdot \sum_{\pi \in \mathcal{S}_n} \OPT(H^\pi)} \,. \]
		Now, it holds that
		\begin{align*}
		\frac{\E[\A(I_n(F))]}{\E[\OPT(I_n(F))]}
		&= \frac{\sum_{H \in \mathcal{H}} \left(\lambda_H \cdot \sum_{\pi \in \mathcal{S}_n} \A(H^\pi)\right) }{\sum_{H \in \mathcal{H}} \left(\lambda_H \cdot \sum_{\pi \in \mathcal{S}_n} \OPT(H^\pi)\right)} & \text{Claim~\ref{claim:CommonProbInEquivClass}}\\
		&\leq \frac{\lambda_I \cdot \sum_{\pi \in \mathcal{S}_n} \A(I^\pi)}{\lambda_I \cdot \sum_{\pi \in \mathcal{S}_n} \OPT(I^\pi)} & \text{Claim~\ref{claim:RatioOfSumsUB}}\\
		&= \frac{\frac{1}{n!} \cdot \sum_{\pi \in \mathcal{S}_n} \A(I^\pi)}{\frac{1}{n!} \cdot \sum_{\pi \in \mathcal{S}_n} \OPT(I^\pi)} \\
		&= \frac{\E[\A(I^\sigma)]}{\OPT(I)} \,.
		\end{align*}
		The second claim follows immediately from $\OPT(I) \geq \sum_{i=1}^{n} X_i \geq cn$.
	\end{proof}
\noindent
It remains to show \MakeUppercase Claims\nobreakspace \ref {claim:CommonProbInEquivClass} and\nobreakspace  \ref {claim:RatioOfSumsUB}.

	\begin{proof}[Proof of \MakeUppercase Claim\nobreakspace \ref {claim:CommonProbInEquivClass}]
		We first show that all lists of the same equivalence class have the same probability to be an outcome of $I_n(F)=(X_1,\ldots,X_n)$.
		Let $I_1=(a_1,\ldots,a_n)$ and $I_2=(b_1,\ldots,b_n)$ with $I_1 \sim I_2$. Note that both lists contain the same multiset of items. Let $m$ be the number of different item sizes and let $z_i$ be the multiplicity of size $y_i$ for $i \in [m]$ in $I_1$ (and thus in $I_2$).
		As the random variables $X_i$ are drawn i.i.d., we have
		\begin{multline*}
		\Pr[I_n(F) = I_1] = \prod_{i=1}^{n} \Pr[X_i = a_i] = \prod_{j=1}^{m} \left(\Pr[X_1 = y_j]\right)^{z_j}
		= \prod_{i=1}^{n} \Pr[X_i = b_i] = \Pr[I_n(F) = I_2] \,.
		\end{multline*}

		Now, consider any equivalence class $[H]$ for $H \in \mathcal{H}$. All lists $I \in [H]$ have the same probability $\Pr[I_n(F) = I] =: \lambda_H$ to be an outcome of $I_n(F)$.
		Further, we have $[H] = \{ H^\pi \mid \pi \in \mathcal{S}_n \}$ and thus
		\begin{multline*}
		\E[\ALG(I_n(F))]
		= \sum_{I \in \mathcal{I}} \ALG(I) \cdot \Pr[I_n(F)=I] 
		= \sum_{H \in \mathcal{H}} \sum_{I \in [H]} \ALG(I) \cdot \Pr[I_n(F)=I] \\
		= \sum_{H \in \mathcal{H}} \left(\lambda_H \cdot \sum_{I \in [H]} \ALG(I)\right)
		= \sum_{H \in \mathcal{H}} \left(\lambda_H \cdot \sum_{\pi \in \mathcal{S}_n} \ALG(I^\pi) \right) \,.
		\end{multline*}
	\end{proof}

\MakeUppercase Claim\nobreakspace \ref {claim:RatioOfSumsUB} is easy to see:
	\begin{proof}[Proof of \MakeUppercase Claim\nobreakspace \ref {claim:RatioOfSumsUB}]
		Define $M := \max_{j \in [m]} (a_j / b_j)$. We have
		\[
		\sum_{j=1}^{m} a_j
		= \sum_{j=1}^{m} \frac{a_j}{b_j} \cdot b_j
		\leq \sum_{j=1}^{m} M b_j
		= M \cdot \sum_{j=1}^{m} b_j \,.
		\] 
	\end{proof}

\subsection{Markov chain from Figure~\ref{fig:markovChain}}
\label{app:stationaryDistribution}

\begin{lemma}
	The Markov chain from Figure 3 converges to the stationary distribution $\omega = (\omega_\mathsf{A},\ldots,\omega_\mathsf{I})$ with
	\[
	\omega = \frac{1}{\lambda}
	\left( 1, ~ p, ~ q+pq\vartheta, ~ p^2, ~ 2pq+p^2q\vartheta, ~ q^2+2pq^2\vartheta, ~ \vartheta, ~ q\vartheta, ~ q^2 \vartheta \right) \,,
	\]
	where 
	$\vartheta=\frac{p^3}{1-q^3}$ and  $\lambda= \vartheta q \left( 3 - q^2 \right) + \vartheta + 3$.
\end{lemma}
\begin{proof}
	As the Markov chain from Figure\nobreakspace \ref {fig:markovChain} is irreducible and aperiodic, it converges to a unique stationary distribution. Let $\omega=(\omega_\mathsf{A},\ldots,\omega_\mathsf{I})$ be this distribution, then we have the following system of equations. 
	\begin{align}
	\omega_\mathsf{A} &= \omega_\mathsf{E} + \omega_\mathsf{F} + p \omega_\mathsf{G} + q \omega_\mathsf{D} 					\tag{Q1} \label{eq:statDistA}\\
	\omega_\mathsf{B} &= p\omega_\mathsf{A} 											\tag{Q2} \label{eq:statDistB}\\
	\omega_\mathsf{C} &= q\omega_\mathsf{A} + p\omega_\mathsf{H} 									\tag{Q3} \label{eq:statDistC}\\
	\omega_\mathsf{D} &= p\omega_\mathsf{B} 											\tag{Q4} \label{eq:statDistD}\\
	\omega_\mathsf{E} &= q\omega_\mathsf{B} + p\omega_\mathsf{C} 									\tag{Q5}\label{eq:statDistE}\\
	\omega_\mathsf{F} &= q\omega_\mathsf{C} + p\omega_\mathsf{I} 									\tag{Q6}\label{eq:statDistF}\\
	\omega_\mathsf{G} &= p\omega_\mathsf{D} + q\omega_\mathsf{I} 									\tag{Q7}\label{eq:statDistG}\\
	\omega_\mathsf{H} &= q\omega_\mathsf{G} 											\tag{Q8}\label{eq:statDistH}\\
	\omega_\mathsf{I} &= q\omega_\mathsf{H} 											\tag{Q9}\label{eq:statDistI}\\
	1 &= \omega_\mathsf{A}+\omega_\mathsf{B}+\omega_\mathsf{C}+\omega_\mathsf{D}+\omega_\mathsf{E}+\omega_\mathsf{F}+\omega_\mathsf{G}+\omega_\mathsf{H}+\omega_\mathsf{I}  \tag{Q10} \label{eq:statDistSum} \,.
	\end{align}
	We define $\vartheta:=\frac{p^3}{1-q^3}$ and  $\lambda:= \vartheta q \left( 3 - q^2 \right) + \vartheta + 3$ and claim that 
	\[
	\omega = 
	\frac{1}{\lambda}
	\left( 1, ~ p, ~ q+pq\vartheta, ~ p^2, ~ 2pq+p^2q\vartheta, ~ q^2+2pq^2\vartheta, ~ \vartheta, ~ q\vartheta, ~ q^2 \vartheta \right)
	\]
	satisfies (\ref{eq:statDistA}) to (\ref{eq:statDistSum}). 
	In fact, the validity of equations (\ref{eq:statDistB}) to (\ref{eq:statDistI}) can be seen immediately.
	To verify (\ref{eq:statDistA}) and (\ref{eq:statDistSum}), we first observe that for any $i \geq 0$ and $1 \leq j \leq 3$, it holds that
	\begin{equation}
	\label{eq:binomialFormula}
	p^i q^j = 
	\begin{cases}
	p^i - p^{i+1} & j=1 \\
	p^i - 2 p^{i+1} + p^{i+2} & j=2 \\	
	p^i - 3 p^{i+1} + 3 p^{i+2} - p^{i+3}& j=3 \,. \\		
	\end{cases}
	\end{equation}
	Now, we prove (\ref{eq:statDistA}). We have
	\begin{align*}
	& \quad \lambda \cdot \left( \omega_{\mathsf{E}} + \omega_{\mathsf{F}} + p \omega_{\mathsf{G}} + q \omega_{\mathsf{D}} \right) \\
	&= (2pq + p^2 q \vartheta ) + (q^2 + 2pq^2 \vartheta) + p \vartheta + qp^2 \\
	&= \vartheta \cdot \left( p^2q + 2pq^2 + p \right) + 2pq + q^2 + p^2q \\
	\overset{\text{eq.(\ref{eq:binomialFormula})}}&{=}  \vartheta \cdot \left( p^2 - p^3 + 2 \cdot (p - 2p^2 + p^3) + p \right)
	+ 2 \cdot (p-p^2) + (1 - 2p + p^2) + (p^2 - p^3) \\
	&= \vartheta (3p - 3p^2 + p^3) + 1 - p^3 \\
	&= 1 \\
	&= \lambda \cdot \omega_\mathsf{A} \,,
	\end{align*}
	which implies (\ref{eq:statDistA}). Equation (\ref{eq:statDistSum}) follows from the following calculation.
	\begin{align*}	
	&\quad \lambda \cdot \left( \omega_\mathsf{A}+\omega_\mathsf{B}+\omega_\mathsf{C}+\omega_\mathsf{D}+\omega_\mathsf{E}+\omega_\mathsf{F}+\omega_\mathsf{G}+\omega_\mathsf{H}+\omega_\mathsf{I}  \right) \\
	&= 1 + p + (q+pq\vartheta) + p^2 + (2pq+p^2q\vartheta) + (q^2+2pq^2\vartheta) + \vartheta + q\vartheta + q^2 \vartheta \\
	&= q \vartheta\cdot \left( p + p^2 + 2pq + 1 + q \right) + (1 + p + q + p^2 + 2pq + q^2) + \vartheta \\
	\overset{\text{eq.(\ref{eq:binomialFormula})}}&{=}
	q \vartheta\cdot \left( p + p^2 + 2 \cdot (p - p^2) + 1 + 1 - p \right) \\
	&~~~~~~ + (1 + p + 1 - p + p^2 + 2 \cdot (p-p^2) + 1 - 2p + p^2) + \vartheta \\
	&= q \vartheta\cdot \left( 3 - q^2 \right) + 3 + \vartheta \\
	&= \lambda \,.
	\end{align*}
	%
	%

\end{proof}

\end{document}